\documentclass{aa}
\usepackage{graphicx}
\usepackage{txfonts}
\usepackage{natbib}

\newcommand{\kev}{keV}
\newcommand{\cm}{cm$^{-3}$}

\newcommand{\etal}{et al.}
\newcommand{\mcg}{MCG--6-30-15}
\newcommand{\iras}{IRAS~13349+2438}

\begin{document}

%\headnote{Research Note}
\title{On the Location and Composition of the Dust in the \mcg\ Warm Absorber}
\titlerunning{Dust in the MCG--6-30-15 warm absorber}

\author{D. R. Ballantyne, Joseph C. Weingartner and
N. Murray\thanks{Canada Research Chair in Astrophysics}}
\authorrunning{Ballantyne et al.}

\offprints{D. Ballantyne}

\institute{Canadian Institute for Theoretical Astrophysics, McLennan Labs,
60 St. George Street, Toronto, Ontario, Canada M5S 3H8\\
\email{ballantyne, weingart, murray@cita.utoronto.ca}}

\date{}

\abstract{
The warm absorber observed in the Seyfert 1 galaxy \mcg\ is known to
consist of at least two zones and very likely contains dust.
\textit{Hubble Space Telescope} images of \mcg\ show a dust lane
crossing the galaxy just below the nucleus. In this paper, we argue
that this dust lane is responsible for the observed reddening of the
nuclear emission and the Fe~\textsc{i} edge hinted at in the
\textit{Chandra} spectrum of \mcg. We further suggest that the gas
within the dust lane can comprise much of the low ionization component
(i.e., the one contributing the O~\textsc{vii} edge) of the observed warm
absorber. Moreover, placing the warm absorbing material at such
distances (hundreds of pc) can account for the small outflow velocities
of the low ionization absorption lines as well as the constancy of the
O~\textsc{vii} edge. Photoionization models of a dusty interstellar gas cloud
(with a column appropriate for the reddening toward \mcg) using a toy
Seyfert~1 spectral energy distribution show that it is possible to
obtain a significant O~\textsc{vii} edge ($\tau \sim 0.2$) if the material is
$\sim$150~pc from the ionizing source. For \mcg, such a distance is
consistent with the observed dust lane. We emphasize the point first
made by Kraemer \etal: dusty interstellar material will likely
contribute to the warm absorber, and should be included in spectral
modeling.

The current data on \mcg\ is unable to constrain the dust composition
within the warm absorber. Astronomical silicate is a viable candidate,
but there are indications of a very low O abundance in the dust, which
is inconsistent with a silicate origin. If true, this may indicate
that there were repeated cycles of grain destruction and growth from
shocks in the interstellar medium of \mcg. Pure iron grains are an
unlikely dust constituent due to the limit on their abundance in the
Galaxy, yet they cannot be ruled out. The high column densities
inferred from the highly ionized zone of the warm absorber implies
that this gas is dust-free.
}

\maketitle

\section{Introduction}
\label{sect:intro}
Soft X-ray absorption by photoionized gas was first used by
\citet{hal84} and \citet{psp90} to explain the unusual spectrum of
the quasar MR~2251--178.  This ``warm absorber'' was characterized by
\textit{ROSAT} spectra of Seyfert~1s which showed absorption edges due
to O~\textsc{vii} and O~\textsc{viii}
\citep{np92,nan93,fio93,tur93}. Warm absorber studies matured with
the launch of the more sensitive \textit{ASCA} observatory, which
allowed more detailed spectral modeling
\citep[e.g.,][]{fab94,gtn95,rf95}. Analysis of a large sample of
Seyfert~1s observed by \textit{ASCA} showed that approximately 50\%
exhibit absorption features from photoionized gas in their soft X-ray
spectra \citep{rey97,geo98}. Currently, the study of the warm
absorber is being revolutionized with observations from the dispersion
gratings onboard \textit{Chandra} and \textit{XMM-Newton} which have
been able to resolve individual absorption lines from a myriad of
metals and charge states in the warm gas
\citep[e.g.,][]{kaa00,kas01,col01,lee01,kas02}. Photoionization
modeling has then shown that, in many cases, more than one ionization
parameter\footnote{There are some objects which can still be modeled
adequately with only one ionization parameter, e.g., Mrk~509
\citep{yaq03}.} is needed to describe the observed spectrum
\citep[e.g,][]{mfr00,kas01,kaa02}. Also, the positions of the lines
point toward outflow velocities on the order of a few hundred to a few
thousand~km~s$^{-1}$. Thus, this warm absorbing gas seems to be in the
form of an outflowing wind \citep[cf.,][]{bks00,elv00}. However,
there remains a major uncertainty in the location of the gas, with
many models considering the broad-line region
\citep[e.g.,][]{rf95,net96,geo98} or the putative obscuring torus
\citep[e.g.,][]{kk95,kk01} as the most likely origin for the warm
absorber.

A possible constraint on the location of at least some of this warm
gas may be provided if it contains any dust, which sublimates at the
radius of the broad-line region \citep{bar87} for a typical active
galactic nucleus (AGN). Dusty warm absorbers (hereafter, DWA) were
first considered for the quasar \iras\ \citep{bfp96,skb99} and the
Seyfert~1 galaxy \mcg\ \citep{rwfc97}. In both of these AGN, the
column of neutral H inferred from the reddening is significantly
larger than that inferred from the neutral absorption in the soft
X-ray band, but is of the same order as the column of ionized gas
inferred from the warm absorber \citep{rwfc97}.  This suggests that
the dust is similar to Galactic dust and resides within the warm
ionized gas, and may significantly affect the observed soft X-ray
spectrum (\citet{kf97} and references therein;
\citealt{kb98}). Spectroscopic evidence for dust in X-ray warm
absorbers has now been found by \citet{lee01} in the \textit{Chandra}
gratings observation of \mcg\ (this seems to have been confirmed by
the very recent \textit{XMM-Newton} data of \citealt{tur03}). This
high resolution spectrum exhibited a sharp drop\footnote{An
alternative explanation for this drop (based on \textit{XMM-Newton}
data) is that it is the blue edge of a relativistically broadened
O~\textsc{viii} Ly$\alpha$ line \citep{bran01,sak02}. However, this
interpretation has been challenged from both observational
\citep{lee01} and theoretical \citep{brf02} points of view.} at
$\sim$0.7~\kev\, consistent with the L3 absorption edge from neutral
Fe. A similar feature was also found in a \textit{Chandra} spectrum of
the Galactic X-ray binary Cyg~X-1 \citep{sch02}. The column implied by
the depth of the Fe edge in \mcg\ is of the right order to explain the
observed reddening ($A_V \approx \tau_V \approx 3E(B-V) = 1.8$;
\citealt{rwfc97}), assuming Galactic-type dust located within the warm
absorber.

Rather than place the DWA near the central engine of the AGN, \citet{kra00}
(see also \citealt{ck01}) argued for the existence of a ``lukewarm absorber''
outside the narrow-line region. This gas would have sufficient column to
explain the observed reddening, and has been ionized to the point
where hydrogen is fully stripped, but the metals would only be moderately
ionized and would exhibit strong UV absorption lines rather than
O~\textsc{vii} or O~\textsc{viii} edges. Thus, this model requires an inner warm
absorber to account for the highly ionized oxygen features
\citep{kra00}. The lukewarm absorber has been shown to be
consistent with the X-ray \citep{kra00} and UV \citep{cren01} data of
NGC~3227, as well as the UV spectrum of Ark~564 \citep{cren02}.

In the case of \mcg\ ($z=0.008$, $L_{\mathrm{2-10 \, keV}} \approx
10^{43}$~erg~s$^{-1}$; see Table~\ref{tab:columns} for a summary of
the absorbing columns), it was clear from early \textit{ASCA}
variability studies that a multi-zone warm absorber was needed
\citep[e.g.,][]{ot96,mfr00}. 
\begin{table}
\caption{A summary of the absorbing columns toward \mcg. Reference 1 =
\citet{ewl89}, 2 = \citet{rwfc97}, 3 = \citet{lee02b}.}
\begin{tabular}{ccccc}
Type & $N_{\mathrm{H}}$ (cm$^{-2}$) & Derived from & Origin & Ref. \\ \hline
Cold & $4.06 \times 10^{20}$  & soft X-ray cutoff & Galaxy &
1 \\
?? & $4 \times 10^{21}$ & reddening & intrinsic & 2 \\
Warm & $\sim 10^{23}$ & soft X-ray edges and lines & intrinsic &
3 \\
\end{tabular}
\label{tab:columns}
\end{table}
In particular, the O~\textsc{viii} edge was
found to anticorrelate with the source luminosity while the O~\textsc{vii}
edge seemed to remain constant \citep{ot96,orr97}. The
\textit{Chandra} observations found a strong O~\textsc{vii} edge
($\tau_{\mathrm{O VII}} \sim 0.7$) and a series of O~\textsc{vii} absorption
lines within 200~km~s$^{-1}$ of the systemic velocity of \mcg\
\citep{lee01}. An updated analysis presented by \citet{lee02b} showed
that many of the other low-ionization absorption lines also have very small
outflow velocities. These data point to a distribution (in
space/velocity) of ionized absorbers, consistent with the earlier
\textit{ASCA} results.

\begin{figure}
\resizebox{\hsize}{!}{\includegraphics{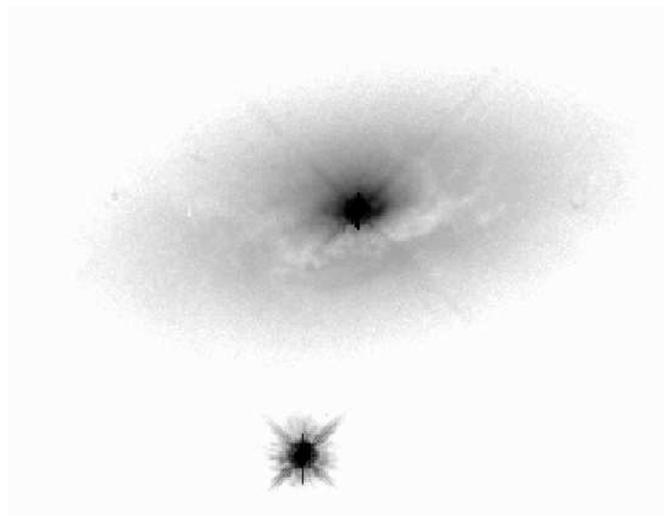}}
\caption{\textit{Hubble Space Telescope} image of \mcg\ \citep{mgt98}. A dust
lane is apparent crossing the southern part of the galactic disk.}
\label{fig:hst}
\end{figure}

A \textit{Hubble Space Telescope} (\textit{HST}) image of \mcg\
shows a distinct dust lane that cuts across the southern part of the
galactic disk (Fig.~\ref{fig:hst}; \citealt{mgt98}; see also \citealt{fwm00}). 
This material is likely intrinsic to the galaxy and not a
foreground effect such as Galactic cirrus (e.g., M81;
\citealt{ar87}).  Therefore, in Sect.~\ref{sect:mcg}, we follow the
ideas of \citet{kra00}, but argue that at least part of the low
ionization warm absorber and all of the observed reddening in \mcg\
originates within this dust lane. It is not our intention to construct
a detailed spectral model of the DWA in \mcg; rather we use this
system as an example of how material in the dusty interstellar medium
(ISM) of the host galaxy can have a significant observational impact
on the study of absorption in AGN spectra.

The composition of the dust in the \mcg\ DWA is also a puzzle. For Cyg~X-1,
where the high signal-to-noise allowed a comparison to laboratory data,
\citet{sch02} claimed that pure metallic Fe was the best fit to the
edge structure. This is surprising, since the expected microwave
emission (at $\approx$ 90~GHz) from such grains is not observed in the
Galaxy \citep{dl99}. \citet{lee01} argued that the Cyg~X-1 iron edge was
similar to the one observed in \mcg, and thus pure iron grains may be
present in the DWA. In Sect.~\ref{sect:dust} we discuss the
constraints on a pure iron component in astronomical dust, and
consider if it is a plausible composition for the dust in the \mcg\
DWA.

\section{The Location of the DWA}
\label{sect:mcg}
The dust lane seen in Fig.~\ref{fig:hst} passes just south of the
nucleus, so, judging from the Galactic distribution \citep{dl90}, our
line of sight likely passes through a column of a few times
10$^{21}$~cm$^{-2}$. This is of the same order as that needed to
provide the observed reddening in \mcg\, assuming Galactic-type grains
($N_{\mathrm{H}} \approx 4 \times 10^{21}$~cm$^{-2}$;
\citealt{rwfc97}). Further evidence for a low density origin of the DWA is the
recombination timescale argument of \citet{ot96}. They noted that the
constancy of the O~\textsc{vii} edge in the long 1994 \textit{ASCA}
observation of \mcg\ implies $n_{\rm H} < 2 \times 10^5 \, {\rm
cm}^{-3}$ in the outer absorber. ISM gas with a fixed hydrogen number
density of $n_\mathrm{H} = 10$~\cm\ will be used as the canonical
absorber for the remainder of the paper.

A simple photoionization argument can be made to constrain the
distance of the O~\textsc{vii} absorber. We define the following ionization
parameter
\begin{equation}
U_{\mathrm{H}} \equiv {L_> \over 4 \pi r^2 c n_{\rm H}
\left<h\nu\right>},
\label{eq:ip}
\end{equation}
where $L_>$ is the AGN luminosity beyond the Lyman edge,
$n_{\mathrm{H}}$ is the total hydrogen number density, and
$\left<h\nu\right>$ is an average ionizing photon energy.
Photoionization equilibrium gives
$\xi_{\mathrm{H}}=n_{\mathrm{H^0}}/n_{\mathrm{H}} =
(\alpha_{\mathrm{H}}(T)/a_{\mathrm{H}}c)U_{\mathrm{H}}^{-1}$, where
$n_{\mathrm{H^0}}$ is the number density of neutral H,
$\alpha_{\mathrm{H}}$ is the recombination coefficient of H at
temperature $T$ and $a_{\mathrm{H}}$ is the photoionization
cross-section for H. This expression is valid only when $\xi \ll
1$. Generalizing for a hydrogenic ion of an element with atomic number
$Z$ gives
\begin{equation}
\xi_Z = { \alpha_{\mathrm{H}}(T) \over a_{\mathrm{H}}c }
U_{\mathrm{H}}^{-1} Z^{4+2\beta} = (1.4 \times 10^{-6})
U_{\mathrm{H}}^{-1} Z^{4+2\beta},
\label{eq:xi}
\end{equation}
where we have assumed $L_{\nu} \propto \nu^{-\beta}$ for the ionizing
luminosity ($\beta \approx 1.6$ for radio-quiet quasars;
\citealt{tel02}), and used $a_{\mathrm{H}}=6 \times 10^{-18}$~cm$^{2}$
and $\alpha_{\mathrm{H}}$(20,000~K)=2.5$\times
10^{-13}$~cm$^{3}$~s$^{-1}$ \citep{ost89}. With a typical Seyfert~1
Lyman luminosity of $L_> = 10^{44}$~erg~s$^{-1}$, and assuming
$\left<h\nu\right>=45$~eV, the ionization parameter of a
$n_{\mathrm{H}}=10$~cm$^{-3}$ cloud at 1~kpc is
0.04. Equation~\ref{eq:xi} then gives $\xi_{\mathrm{H}} \approx
10^{-5}$ and $\xi_{\mathrm{O}} \approx 100$, implying that although
hydrogen is still entirely ionized at this distance, oxygen has
started to recombine. If we now consider gas only 10~pc from the AGN,
the values of $\xi$ will drop by 10$^4$, and O is predominantly
fully-stripped. Therefore, the O~\textsc{vii} absorber must lie between 10 and
1000~pc. This calculation assumes photoionization equilibrium and is
therefore valid only on timescales longer than the recombination time,
$t_{\mathrm{rec}} \sim 1/n_e \alpha_{\mathrm{H}}(T) \sim 10^4$~years,
where $n_e$ is the free electron density. The radial range derived
above will be reasonable as long as the luminosity of the central
engine did not change by factors greater than $\sim 10$ over
$t_{\mathrm{rec}}$. To more accurately determine the properties of an
ISM warm absorber, numerical photoionization models must be employed.

We used Cloudy~96Beta4 \citep{fer02} to predict the ionization
structure of a $n_\mathrm{H} = 10$~\cm\ cloud of ISM gas at various
distances from an AGN. The neutral hydrogen column density was fixed
at $N_{\rm H} = 4 \times 10^{21}$~cm$^{-2}$, which, for a Galactic
gas-to-dust ratio, is the minimum needed to provide the reddening toward \mcg\
($N_{\mathrm{H}}=4$--7$\times 10^{21}$~cm$^{-2}$;
\citealt{rwfc97}). The dust (silicate plus graphite) and gas-phase
metal abundances were fixed at the ISM values described by
\citet{fer02}. The cloud was illuminated with a
``standard'' AGN continuum (see pg.~34 in \citealt{fer02}) with
$\alpha_{\mathrm{ox}}=-1.4$ and a X-ray power-law photon-index
$\Gamma=2$. The ``big blue bump'' was characterized by a temperature
of $1.4\times 10^5$~K (the maximum \citet{ss73} accretion disc
temperature for a 10$^7$~M$_{\sun}$ black hole accreting at 0.1 of its
Eddington rate) and a UV slope of $\alpha\mathrm{(UV)}=-0.5$. This
spectral energy distribution (SED) is not intended to be a realistic
model of the \mcg\ continuum (which is unknown because of the large
reddening), but rather representative of a generic AGN. The
normalization of the SED was set by defining the 2--10~\kev\
luminosity to be 10$^{43}$~erg~s$^{-1}$, typical of many Seyfert~1s
\citep{rey97}.

Models were calculated with the inner edge of the gas cloud at various distances $r$
from the continuum source, and the computed O~\textsc{vii} column density was
compared with the result from the \textit{Chandra} observation of \mcg,
$N_{\mathrm{O VII}} \approx 2.5\times 10^{18}$~cm$^{-2}$
\citep{lee01}.  The maximum O~\textsc{vii} column found in the Cloudy runs
was $6.1\times 10^{17}$~cm$^{-2}$ for a cloud distance of
150--175~pc. Models with the gas closer in were too ionized, and if
the cloud was further out, it was not ionized enough. Of course,
lowering the density by a factor $f$ would allow a distance $\sqrt{f}$
larger.  However, if the distance exceeds $\approx \,$200--250 pc,
then, even with $n_{\rm H} = 10 \, {\rm cm}^{-3}$, the physical length
of the column exceeds the distance to the AGN. 
A similar calculation with a power-law SED of energy index $-1.17$
between 0.013 and 100~\kev\ resulted in $N_{\mathrm{O VII}} =
6.2\times 10^{17}$~cm$^{-2}$ at $r=150$~pc. While these models cannot
account for the entire O~\textsc{vii} column in \mcg\ (it is likely that some
fraction of the O~\textsc{vii} edge originates within the inner warm
absorber), we have shown that this dusty ISM cloud will have an
detectable impact on the observed spectrum (see also \citealt{kb98}).

\begin{figure}
\resizebox{\hsize}{!}{\includegraphics{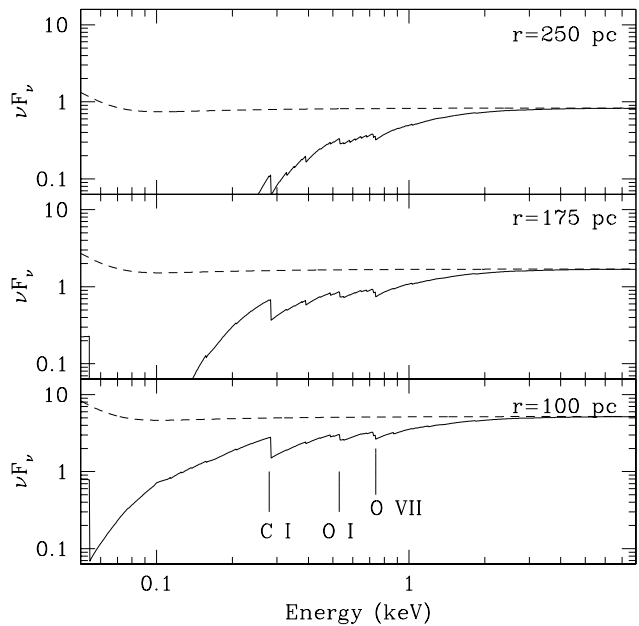}}
\caption{The results of photoionization models of a DWA shown over the
X-ray band. Constant density ($n_{\mathrm{H}} = 10$~\cm) gas with $N_{\mathrm{H}}=4\times 10^{21}$~cm$^{-2}$ was
illuminated by a toy AGN SED (shown with the dashed line), and the
transmitted spectrum is shown with the solid line. Typical
interstellar dust was included at Galactic abundances. The
positions of the most prominent photoelectric absorption edges are
indicated in the lower panel. The gas was placed at different
distances $r$ from the AGN. The greatest column of O~\textsc{vii}
($N_{\mathrm{O VII}}=6\times 10^{17}$cm$^{-2}$) was found when
$r=$150--175~pc.}
\label{fig:cloudy}
\end{figure}

To illustrate the extent of this impact, the incident and transmitted
continua for the AGN models with $r=100$, $175$, and $250$~pc are
shown in Fig.~\ref{fig:cloudy}.
These runs have
$U_{\mathrm{H}}=3.7$, $1.2$ and $0.6$, respectively.  The transmitted
spectra show significant warm absorption in the soft X-ray band with
strong edges at 0.28, 0.53 and 0.74~\kev. These first two are due to
C~\textsc{i} and O~\textsc{i} in the grains (we include both graphite and silicate
dust), while the latter is the O~\textsc{vii} edge. Of the three models shown
here, the depth of the O~\textsc{vii} edge is greatest when $r=175$~pc, with
$\tau_{\mathrm{O VII}} = 0.15$. The Fe~\textsc{i} edge from the dust is
noticeable as a small notch just redward of the O~\textsc{vii} edge, as was
observed by \citet{lee01}. The majority of the opacity below 0.3~\kev\
is due to absorption by He~\textsc{ii}, as was found in the lukewarm absorber
of \citet{kra00}. Therefore, such absorption will need to be included
in absorption models for sources such as \mcg\ where distant material
is contributing to the warm absorber. Indeed, in their multi-zone
DWA fit to the new \textit{XMM-Newton} data, \citet{tur03} require a
low-ionization component to account for additional soft X-ray
absorption that is inconsistent with cold absorption. It is quite
likely that this excess attenuation is caused by He~\textsc{ii} absorption.

The fact that there is a significant column of He~\textsc{ii} ($5.9 \times
10^{18}$~cm$^{-2}$) in the models shows that the gas is very close to
recombining to a more neutral configuration. This is in large part due
to the presence of dust which, for a Galactic size distribution,
dominates the continuum opacity in the extreme UV (EUV) and soft
X-ray bands (see Fig.~\ref{fig:grains}). Indeed, our Cloudy models show that
neutral oxygen begins to be present in the gas when $r=175$~pc. Thus,
our results seem to be fine-tuned in the sense that the greatest
O~\textsc{vii} column is found when the gas is on the verge of
recombining. This argues that it may be difficult, in general, for ISM
gas to contribute to the warm absorber.

In this section we have argued that ISM material associated with dust
lanes in the host galaxy of an AGN is a natural contributor to the
observed DWAs. Cloudy models suggest that this gas, when illuminated
by a toy AGN SED, can produce significant warm absorption (including a
significant O~\textsc{vii} edge) when placed $\sim$150~pc from the AGN.  This
may be particularly relevant to \mcg, where a dust lane does pass just
below the nucleus.  Using Fig.~\ref{fig:hst} and the plate scale from
\citet{mgt98}, we can estimate the inner edge of the dust lane from
the \textit{HST} image. Interestingly, this is also of the order of
150--450~pc, assuming $H_0=50$~km~s$^{-1}$~Mpc$^{-1}$.

It may be difficult applying this idea to other sources which harbor
DWAs (e.g., \iras). The dust lane in \mcg\ lies nearly along
our line of sight and may be a relatively unusual alignment. Nevertheless,
the strong ionizing power of an AGN will have a great impact on its
local ISM. The observational consequences of these effects will depend
both on the viewing angle into the source \citep{ck01} and on the SED
of the AGN. Indeed, if our line-of-sight into \mcg\ were slightly
different and passed through the dust lane, rather than just above it,
the obscuration could be high enough that the AGN would appear as a
Seyfert~2 \citep[see, e.g.,][]{mat00}.

\section{Dust Properties}
\label{sect:dust}
\subsection{Composition}
\label{sub:comp}

\begin{figure}
\resizebox{\hsize}{!}{\includegraphics{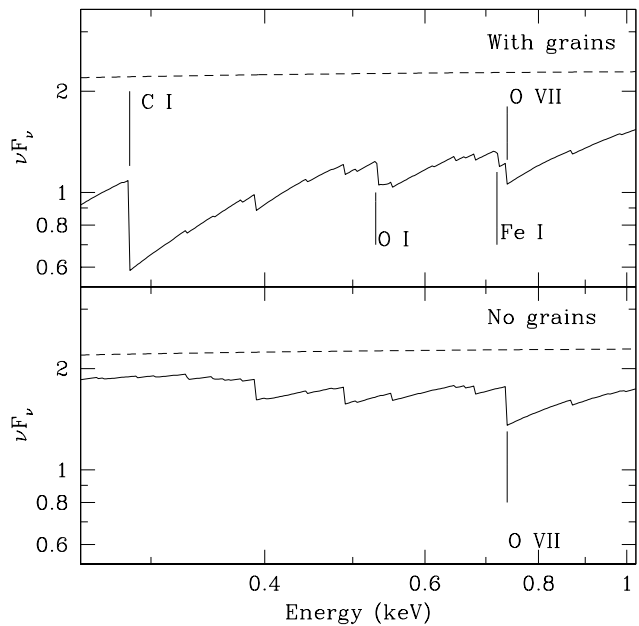}}
\caption{The effect of grains on the transmitted
spectrum (see also \citealt{kb98}). Both panels show the results of
Cloudy models where the gas cloud was 150~pc from the AGN. The dashed
curve shows the incident spectrum and the solid curve shows the
transmitted spectrum. The positions of the C~\textsc{i}, O~\textsc{i},
Fe~\textsc{i}, and O~\textsc{vii} edges are indicated. Solar
abundances were used for the dust-free model, except that the oxygen
abundance was set to the F and G star value from \citet{sm01}.}
\label{fig:grains}
\end{figure}

As the EUV/soft X-ray opacity in the DWA is dominated by dust, the
absorption properties depend on the grain composition.  The top panel
of Fig.~\ref{fig:grains} shows the transmitted spectrum for the
$r=150$~pc model with Galactic silicate and graphite dust abundances
\citep{mrn77}.
 In the bottom panel, the spectrum is
displayed for the case where the elements in the dust have been
returned to the gas. We assumed that the total (gas + dust) abundances
are equal to solar \citep{gs98} except for O, for which we took
O/H=$4.45 \times 10^{-4}$, a value measured from local young F and G
stars \citep{sm01}. This oxygen abundance has the advantage that it
does not require depletions greater than what can be accounted for by
dust. The DWA model exhibits a C~\textsc{i} edge at 0.28~\kev\ of depth
$\tau = 0.62$ from the graphite dust, an O~\textsc{i} edge at 0.53~\kev\ of
depth $\tau = 0.15$, and an Fe~\textsc{i} edge at 0.72~\kev\ of depth $\tau
= 0.09$, both from the silicate dust. The dust-free model also shows
absorption edges, the largest from O~\textsc{vii} at 0.74~\kev\ of depth $\tau
= 0.26$.

Unfortunately, the High Energy Transmission Gratings (HETG)
\textit{Chandra} spectra only extend to an energy of 0.45~keV, so we
have no X-ray evidence for or against the presence of graphite;
however the Low Energy Transmission Gratings (LETG) can reach these low
energies although the instrumental C edge will need to be well
calibrated to measure the intrinsic edge due to graphite. The
Reflection Grating Spectrometer (RGS) onboard \textit{XMM-Newton} also
does not go below 0.3~\kev, although the low-resolution European
Photon Imaging Cameras (EPIC) are able to do so.

As mentioned in \S~\ref{sect:intro}, the strength of the measured
Fe~\textsc{i} edge from the \textit{Chandra} observation of \mcg\ implies
$N_{\rm Fe \, I} \sim 4 \times 10^{17} \, {\rm cm}^{-2}$, consistent
with the observed reddening for Galactic-type dust.  If the Fe really
is incorporated into silicate grains, then we expect $N_{\rm O \, I} >
N_{\rm Fe \, I}$, since a silicate structural unit contains more O
atoms than Fe atoms. For example, the Cloudy models discussed above
assume a MgFeSiO$_4$ stoichiometry and predict an O~\textsc{i} edge that is
roughly the same depth as the Fe~\textsc{i} edge. \citet{lee02b} claimed
that silicates with an O:Fe number ratio of 2:1 rather than 4:1 were
consistent with the \textit{Chandra} data, but also mentioned that the
data were limited by statistics at the position of the O~\textsc{i}
edge. Therefore, if the silicate dust had an O:Fe ratio of 4:1, it
could have been detected by \textit{Chandra}. However, \citet{sak02}
quote an upper limit of $N_{\rm O \, I}=10^{16}$~cm$^{-2}$ from their
\textit{XMM-Newton} RGS observation of \mcg, which has higher
signal-to-noise, but lower resolution, than the \textit{Chandra}
spectrum. These authors also find $N_{\mathrm{Fe \, I}} = 7\times
10^{16}$~cm$^{-2}$ for the column of neutral iron, which implies a
$V$-band optical depth of 0.07 (assuming a silicate grain radius of
0.1~$\mu$m). Therefore, this column could not account for the observed
reddening of \mcg, unless it is provided predominantly by iron-poor
grains.

Further constraints on the mineralogical structure of the dust can, in
principle, be provided by the detailed structure of the Fe~\textsc{i} edge.
\citet{lee01} found that the edge structure in \mcg\ matches very well
that found in Cyg X-1.  \citet{sch02} showed that the Cyg X-1 edge
could be fairly accurately reproduced if the Fe resides in pure Fe
grains (taking experimental data from \citealt{kk00}), but not if the
Fe is in oxides (taking data from \citealt{croc95}). A measurement of
the oxygen column in the dust could not be made for Cyg~X-1, as the
large O~\textsc{i} edge ($\tau > 2$) was dominated by neutral oxygen in the
ISM along the line-of-sight. Therefore, \citet{sch02} concluded that
pure iron grains were most consistent with the data, and \citet{lee01}
present this as a possibility for the dust in \mcg. However, for
Cyg~X-1 metallic Fe dust is probably ruled out since it would produce
excessive thermal magnetic dipole emission at 90~GHz
\citep{dl99}. Draine \& Lazarian suggest that at most $\sim$5\% of the
Galactic interstellar Fe can be in metallic iron grains. Thus, it
seems unlikely that there is a significant column of metallic iron
grains toward Cyg~X-1. Although we have not performed a detailed
spectral fitting, it appears that the observed edge structure in Cyg
X-1 could be due to Fe in silicates, based on the spectrum of an
olivine sample from \citet{ggmc00}.

The above argument cannot be directly applied to \mcg\ as there are no
microwave observations of this galaxy, so the key to uncovering the
dust composition in the DWA is an accurate measurement of other edges
such as O~\textsc{i}, C~\textsc{i}, or Si~\textsc{i} at 1.85~\kev\
($\tau_{\mathrm{Si~I}} < 0.01$ in the Cloudy models). The
\textit{Chandra} data may imply w\"{u}stite (FeO) or hematite
($\alpha$Fe$_2$O$_3$) dust, but the \textit{XMM-Newton} limit by
\citet{sak02} points to little or no oxygen in the grains.

Perhaps the lack of O~\textsc{i} can be explained by selective destruction
of O-bearing grains (e.g., silicates).  Depletion patterns in the
Galaxy suggest that Fe and Si are incorporated into different dust
populations, with the Fe-bearing component significantly longer-lived
than the Si-bearing component \citep[e.g.,][]{t98}.  If the dust in
the DWA in \mcg\ were subjected to repeated episodes of destruction
(in shocks) and re-growth, then the Si-bearing dust component could be
largely removed while the Fe-bearing component remains largely intact.
Such a process would lead to a reduced $E(B-V)/N_{\rm H}$.  However,
the metallicity at $r \sim 150$~pc from the nucleus of \mcg\ is
probably higher than at the location of the Sun in the Galaxy, so that
$E(B-V)/N_{\rm H}$ in the DWA in this scenario could be comparable to
its value in the local ISM of the Galaxy.  One potential difficulty
for this scenario (and for silicate dust models in general) is that
the destruction timescale inferred by \citet{t98} for the Si-bearing
dust is substantially shorter than predicted for silicate grains
\citep{jth96}.  Counter-intuitively, Jones \etal\ also find the
lifetime of pure Fe dust to be shorter than the lifetimes of graphite
or silicate dust, because Fe grains are accelerated to higher speeds
in the shock.  \citet{wd99} suggested the Fe may largely be
incorporated into the carbonaceous dust population rather than the
silicate component.  In this picture, gas-phase Fe atoms rapidly
accrete onto polycylic aromatic hydrocarbon molecules, forming
organometallic ``sandwich'' molecules, which may ultimately coagulate
to form larger grains.  Laboratory Fe L-edge spectra of such compounds
would be very useful.

\subsection{Other implications}
\label{sub:other}

The \textit{Chandra} data also place a constraint on the inner warm
absorbing gas which is responsible for the \ion{O}{8} edge in the
\textit{ASCA} data. From the depth of the high-ionization absorption
lines, \citet{lee02b} found $N_{\rm H} \sim 10^{23} \, {\rm cm}^{-2}$
for this material.  If the Fe~\textsc{i} edge observation were not
available one might argue that the inner
absorber contains large grains that do not redden the optical nuclear
emission. However, since all of the neutral Fe can be accounted for by the
outer absorber, no such Fe-bearing grains exist in the inner absorber,
suggesting that the inner absorber is dust-free.  This, in turn,
suggests than the inner absorber is located within the dust
sublimation radius, unless an alternative mechanism can be found for
depleting the gas of dust at larger distances.

Indeed, if grains are to be found close to the central engine of an
AGN, they must be shielded from the outpouring radiation. The
radiation pressure opacity of dust is $\sim 10^3$ times that of
electrons \citep{ld93}, so within a certain
radius the dust will see a significantly super-Eddington source. The
gravitational and radiation forces will balance when the dust particle
is outside a mass $\sim 1000M_{\mathrm{BH}}$, where $M_{\mathrm{BH}}$
  is the black hole mass of the AGN. As an example, the
inner velocity profiles of S0 galaxies are approximately linear
\citep{ss96,loy98}, implying that $M(r) \propto r^3$. Thus, if at
$r=1$~pc $M(r)=M_{\mathrm{BH}}$, then the balancing radius for the
dust will be $\sim$10~pc, independent of the black hole
mass. Therefore, if a DWA is at 150~pc it will not feel a
significant radiation force, and should not have a large outflow
velocity. In the case of \mcg, this is consistent with the preliminary results from the
\textit{Chandra} data \citep{lee01,lee02b}, and the low velocity
component inferred by \citet{sak02} from the \textit{XMM-Newton} spectrum.

\section{Summary}
\label{sect:summ}
The DWA in \mcg\ has been the subject of two different X-ray gratings
 observations, but the interpretation of the spectra has been
 controversial \citep{bran01,lee01}. However, the optical/UV reddening
 requires dust along the line-of-sight to the AGN. This dust will
 imprint observable Fe, C, O and Si edges on an X-ray spectrum, and
 indeed an Fe~\textsc{i} edge was likely detected in the \textit{Chandra}
 \citep{lee01} and \textit{XMM-Newton} \citep{tur03} data. The gas associated with this dust has a column of
 a few times 10$^{21}$~cm$^{-2}$ and may be responsible for much of
 the low-ionization component of the warm absorber in \mcg\ (that is,
 it produces the O~\textsc{vii} edge).

Motivated by the the dust lane seen in the \textit{HST} image of \mcg,
 we have argued in this paper that DWAs may reside in the ISM of the
 host galaxy. Photoionization modeling shows that a detectable O~\textsc{vii}
 edge can be produced by such material if it is placed $100$--$200$~pc
 away from the AGN and therefore can contribute to the warm absorption
 features. Furthermore, the dust contained within this gas is
 dynamically and thermally stable and can account for any observed
 reddening. Supporting evidence for this interpretation for \mcg\ is
 given by the very small outflow velocity of the low ionization
 absorption lines, the observed constancy of the O~\textsc{vii} edge, and the
 additional absorption required at low energies.

The composition of the dust in DWAs may be constrained by the relative
strength of the Fe~\textsc{i} and O~\textsc{i} edges, or by the detailed
structure around the Fe~\textsc{i} edge, as was done in Cyg~X-1. The present
\mcg\ data is consistent with silicate dust having a O:Fe ratio
of 2:1 or less, which could result from a process of repeated grain
destruction and growth in the ISM. The presence of pure iron grains
seems unlikely given the limit on their abundance from our own
Galaxy. High signal-to-noise data around the O~\textsc{i} edge, or a
measurement of the C~\textsc{i} edge, is needed to fully exploit the
observations in determining the dust composition in distant AGN. 

\begin{acknowledgements}

We thank Gary Ferland for helpful discussions regarding the Cloudy
models, and Julia Lee for comments on a draft of the manuscript.  This
research was supported by the Natural Sciences and Engineering
Research Council of Canada and by the Canada Research Chair
Program. This research has made use of the NASA/IPAC Extragalactic
Database (NED) which is operated by the Jet Propulsion Laboratory,
California Institute of Technology, under contract with the National
Aeronautics and Space Administration.
\end{acknowledgements}

\end{document}